\documentclass[11pt,english]{article}
\usepackage{graphicx}
\usepackage{graphics}
\usepackage{epsfig}
\usepackage{authblk}
\usepackage{subfigure}
\usepackage{wrapfig}
\usepackage[T1]{fontenc}
\usepackage[latin9]{inputenc}
\usepackage{units}
\usepackage{amsmath}
\usepackage{amssymb}
\usepackage{esint}
\usepackage{babel}
\usepackage{float}
\usepackage{empheq}
\usepackage{subfigure}
\usepackage{wrapfig}

\usepackage{xcolor}
\usepackage[normalem]{ulem}

\usepackage{setspace}
 \title{
Investigation of G\"{o}rtler vortices in high-speed boundary layers via an efficient numerical solution to the non-linear boundary region equations}

\author[1]{Omar Es-Sahli\thanks{email: oe83@msstate.edu}}
\author[1]{Adrian Sescu\thanks{email: sescu@ae.msstate.edu}}
\author[2]{Mohammed Afsar}
\author[3]{Yuji Hattori}

\affil[1]{Department of Aerospace Engineering, Mississippi State University, US}
\affil[2]{Department of Mechanical \& Aerospace Engineering, Strathclyde University, UK}
\affil[3]{Institute of Fluid Science, Tohoku University, Japan}

\begin{document}

\date{}

\maketitle

\begin{abstract}

Streamwise vortices and the associated streaks evolve in boundary layers over flat or concave surfaces due to disturbances initiated upstream or triggered by the wall surface. Following the transient growth phase, the fully-developed vortex structures become susceptible to inviscid secondary instabilities resulting in early transition to turbulence via `bursting' processes. In high-speed boundary layers, more complications arise due to compressibility and thermal effects, which become more significant for higher Mach numbers. In this paper, we study G\"{o}rtler vortices developing in high-speed boundary layers using the boundary region equations (BRE) formalism, which we solve using an efficient numerical algorithm. Streaks are excited using a small transpiration velocity at the wall. Our BRE-based algorithm is found to be superior to direct numerical simulation (DNS) and ad-hoc nonlinear parabolized stability equation (PSE) models. BRE solutions are less computationally costly than a full DNS and have a more rigorous theoretical foundation than PSE-based models. For example, the full development of a G\"{o}rtler vortex system in high-speed boundary layers can be predicted in a matter of minutes using a single processor via the BRE approach. This substantial reduction in calculation time is one of the major achievements of this work. We show, among other things, that it allows investigation into feedback control in reasonable total computational times. We investigate the development of the G\"{o}rtler vortex system via the BRE solution with feedback control parametrically at various freestream Mach numbers $M_\infty$ and spanwise separations $\lambda$ of the inflow disturbances.


\end{abstract}



\section{Introduction}

Streaks formation in pre-transitional boundary layer flows over flat or curved surfaces occur when the amplitude of the freestream disturbances is sufficiently large, or the height of the roughness elements of the surface exceeds a certain critical value. The streamwise velocity component exhibits elongated, so-called `streaky', features characterized by adjacent regions of acceleration (high-speed) and deceleration (low-speed) of fluid particles (Kendall \cite{Kendall}, Matsubara \& Alfredsson \cite{Matsubara}, or Landahl \cite{Landahl}). Elongated streaks in the form of streamwise (G\"{o}rtler) vortices also appear inside a boundary layer flow along a concave surface due to the imbalance between radial pressure gradients posed by the wall, and centrifugal forces (e.g., G\"{o}rtler \cite{Gortler}, Hall \cite{hall1,hall2,Hall6}, Swearingen \& Blackwelder \cite{Swearingen}, Malik \& Hussaini \cite{Malik}, Li \& Malik \cite{Li}, Wu et al. \cite{Wu}, Sescu et al. \cite{Sescu2,Sescu3}, Marensi \& Ricco \cite{Marensi2}, Xu et al. \cite{Xu}). For highly curved walls, for example, vortex formation occurs more rapidly and can significantly alter the mean flow causing the laminar flow to breakdown into turbulence. The literature treating and discussing G\"{o}rtler vortices evolving in incompressible boundary layer flows is rich; we encourage the reader to consult the aforementioned studies and many others that are included in their respective references lists.



G\"{o}rtler vortices in compressible boundary layers have been studied for quite a while motivated by numerous engineering applications such as high-speed flow in engine intakes, flows over the concave surface a turbomachinery blades, or flows over the walls of supersonic nozzles, for example. They were studied by using parallel flow theory by Kobayashi and Kohama \cite{Kobayashi}. Non-parallel effects were studied by El-Hady and Verma \cite{El-Hady}, Hall and Fu \cite{Hall4}, and Hall and Malik \cite{Hall5}. Spall and Malik \cite{Spall} further improved the parallel eigenvalue framework by adding initial conditions to the partial differential equations, assuming zero amplitude perturbations in the external boundary layer (see slight modifications of this approach in Wadey \cite{Wadey} or Dando and Seddougui \cite{Dando}). The number of experiments involving G\"{o}rtler vortices developing in compressible boundary layers is not as large as the number of experiments performed in the incompressible regime. Worth to mention are the experiments of De Luca et al. \cite{De_Luca}, Ciolkosz and Spina \cite{Ciolkosz}, or Wang et al. \cite{Wang}.

While the aforementioned theoretical and numerical studies are relatively old, there has been a resurgence of interest in G\"{o}rtler vortices evolving in high-speed boundary layers in recent years. For instance, Li et al. \cite{Li2} studied the linear and nonlinear growth of G\"{o}rtler vortices in hypersonic boundary layers using the parabolized stability equations (PSE), linear stability analysis, and direct numerical simulations (DNS). They identified multiple sets of unstable secondary instability modes and investigated their linear and nonlinear spatial development. DNS was used to explore the onset of transition in order to determine the most important physical mechanisms associated with high-speed boundary layers. Ren and Fu \cite{Ren} conducted a series of numerical computations to investigate the fundamental, subharmonic and detuned secondary instabilities of G\"{o}rtler vortices in high-speed boundary layer flows with a focus on the Mach number effect. They found that the growth rate associated with G\"{o}rtler vortices decreases with the Mach number and contributes to the appearance of the trapped-layer mode in the primary instability. Chen et al. \cite{Chen} employed DNS and linear stability analysis to explore the transition of stationary G\"{o}rtler vortices in high-speed boundary layer flows by exciting the instability using steady blowing and suction on the wall, similar to the disturbance employed in this work. It was shown that, depending on its frequency and wavelength, the first Mack mode can turn into either a varicose or sinuous mode streak instability, while the second Mack mode turns into a varicose mode. 

In Li et al. \cite{Li3}, DNS and linear secondary instability theory were utilized to study G\"{o}rtler vortices and their associated secondary instabilities in the hypersonic boundary layer flow over a cone featuring a concave aft body. It was revealed that the secondary instability is predominantly characterized by sinuous modes that concentrates in the wall-normal internal shear layer in the lower portion of the mushroom shapes. A form of compressible boundary region equations has been used by Viaro and Ricco \cite{Viaro} to study the evolution of G\"{o}rtler vortices in compressible boundary layer flows. They used a previous approach that was derived by Ricco and Wu \cite{Ricco} as an extension of the incompressible approach of Leib et al. \cite{Leib} to the compressible regime. Within this formalism, both the effect of the initial conditions and the boundary condition at the top of the boundary layer are correctly accounted for. They also explained the formation and growth of thermal streaks, which are thought to play a significant role in the secondary instability process (see also Ricco \cite{Ricco1}). Ricco, Tran \& Ye \cite{Ricco2} and Ricco, Shah \& Hicks \cite{Ricco3} further studied the influence of wall heat transfer and wall suction, respectively, on the thermal streaks. Finally, we mention the work of Song et al. \cite{Song}, where the evolution of first and second Mack mode inside a compressible boundary layer involving G\"{o}rtler vortices is studied in detail via DNS and PSE. Nevertheless, DNS falls short on computation efficiency for large Reynolds numbers that are typically of interest, whereas PSE-based approaches remain largely ad-hoc and suffer from convergence problems \cite{Bagheri_2007}. This justifies investigation into a more efficient numerical solution based on a more robust theoretical foundation.

The present paper focuses on investigating the streamwise vortices and the associated G\"{o}rtler vortices that develop in high-speed boundary layers over concave surfaces using the full nonlinear BRE. The BRE represent the high Reynolds number asymptotic limit of the Navier-Stokes (N-S) equations under the assumption that the streamwise wavenumbers of the disturbances are much smaller than those associated with the crossflow disturbances. This set of equations is parabolic in the streamwise direction allowing for the application of a straightforward marching procedure along the streamwise direction. To this end, upstream conditions have to be imposed to start the calculation, and here we accomplish this by a small disturbance at the wall in the form of wall transpiration. At the upstream boundary, we impose mean flow profiles obtained from a similarity solution applied to compressible boundary layer equations (equivalent to the Blasius solution for an incompressible flow). Since a concave wall is considered in the analysis, the disturbances take the form of G\"{o}rtler instabilities featuring counter rotating pairs of vortices and associated streaks, with streamwise velocity contours plotted in crossflow sections resembling mushroom shapes. We analyze and quantify the evolution of these streaks via contour plots of velocity and temperature in crossflow planes as well as vortex energy, wall shear stress, and wall heat flow distributions versus the streamwise coordinate. We emphasize that the numerical algorithm that is applied to BRE is very efficient, making it suitable for timely parametric studies.

The rest of the paper is organized as follows: In section II, we introduce and describe the mathematical model as we discuss the scaling of various independent/dependent variables, the appropriate initial and boundary conditions, as well as the numerical algorithm. In section III, we report and discuss results for various freestream Mach numbers and spanwise separations, in supersonic and hypersonic regimes. Section IV includes concluding remarks.

\section{Problem formulation and numerical algorithm}

\begin{figure}[H]
 \begin{center}
    \includegraphics[width=12cm]{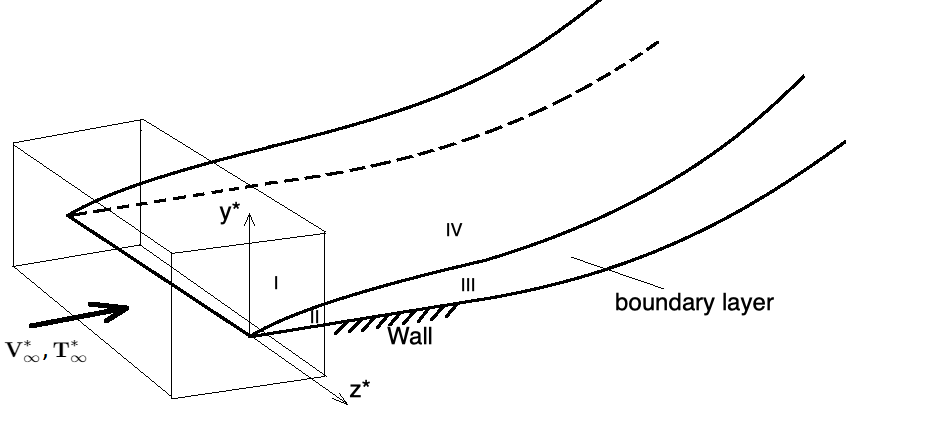}
 \end{center}
  \caption{\label{fig1} The flow domains illustrating the asymptotic structure and the four regions of the flow domain. The box distinguishes regions I and II with I inside the box and outside the boundary layer and II inside the box and inside the boundary layer.}
  \label{f1}
\end{figure}

We consider a compressible flow of uniform velocity $V_{\infty}^*$ and temperature $T_{\infty}^*$ over a curved surface. The air is treated as a perfect gas such that the speed of sound $c_{\infty}^* =\sqrt{\gamma R T_{\infty}^*}$, where $\gamma$ = 1.4
is the ratio of the specific heats, and $R = 287.05 N m/(kg K)$ is the universal gas constant. The Mach number, $M_\infty=V_{\infty}^*/c_{\infty}^*$, is assumed to be of order one. Note that the superscript * and the subscript $\infty$ symbols indicate dimensional freestream quantities. The flow is divided into four regions as in Leib et al. \cite{Leib}, Ricco \& Wu \cite{Ricco}, and Marensi et al. \cite{Marensi} (see figure \ref{f1}). Region I is in proximity to the the leading edge, outside of the boundary layer; the flow is assumed inviscid and the disturbances are treated as small perturbations of the base flow. Region II is the boundary layer in the vicinity of the leading edge with thickness much smaller than the spanwise separation associated with the freestream disturbances; the disturbances are governed by the linearized boundary region equations and the diffusion in the spanwise direction is of the same order of magnitude as that in the wall-normal direction. Region III is the viscous region that follows downstream region II; the boundary layer thickness is of the same order of magnitude as the spanwise separation and the flow is governed by the BRE derived from the full N-S equations by neglecting the streamwise pressure-gradient and the streamwise viscous diffusion (since they evolve on a slow streamwise scale consistent with the vortex structure being elongated in that direction). Region IV, outside the boundary layer above region III, is assumed inviscid since the viscous effects are negligible; the flow is influenced at leading order by the displacement effect due to the increased thickness of the viscous layer underneath it.

The focus in this paper is on region III, where the streamwise wavenumber of disturbances are expected to be small, as previous incompressible G\"{o}rtler flow studies suggest (e.g. Wu et al \cite{Wu}, Marensi \& Ricco \cite{Marensi2}), and the flow is governed by the compressible boundary region equations.

\subsection{Scalings}

All dimensional spatial coordinates $(x^*,y^*,z^*)$ are normalized by the spanwise separation $\lambda^*$ of the disturbances, while the dependent variables by their respective freestream values, and the pressure field is normalized by the dynamic pressure:
\begin{eqnarray}\label{NS}
&&  \bar{t} = \frac{t^*}{\lambda^*/V_{\infty}^*}; \hspace{4mm} \bar{x} = \frac{x^*}{\lambda^*}; \hspace{4mm}
\bar{y} = \frac{y^*}{\lambda^*}; \hspace{4mm} \bar{z} = \frac{z^*}{\lambda^*};  \hspace{4mm}
\bar{u} = \frac{u^*}{V_{\infty}^*}; \hspace{4mm} \bar{v} = \frac{v^*}{V_{\infty}^*}; \hspace{4mm}
\bar{w} = \frac{w^*}{V_{\infty}^*}; \nonumber  \\
&& \bar{\rho} = \frac{\rho^*}{\rho_{\infty}^*}; \hspace{4mm} 
\bar{p} =  \frac{p^*-p_{\infty}^*}{\rho_{\infty}^*V_{\infty}^{*2}}; \hspace{4mm} 
\bar{T} = \frac{T^*}{T_{\infty}^*}; 
\hspace{4mm} \bar{\mu} = \frac{\mu^*}{\mu_{\infty}^*}; \hspace{4mm} \bar{k} = \frac{k^*}{k_{\infty}^*},
\end{eqnarray}
where $(u^*,v^*,w^*)$ are the velocity components, $\rho^*$ the density, $p^*$ is pressure, $T^*$ temperature, $\mu^*$ dynamic viscosity, and $k^*$ thermal conductivity. All quantities with $\infty$ at the subscript represent conditions at infinity.

Reynolds number based on the spanwise separation, Mach number and Prandtl number are defined as
\begin{eqnarray}\label{NS}
R_{\lambda} = \frac{\rho_{\infty}^* V_{\infty}^* \lambda^*}{\mu_{\infty}^*}, \hspace{5mm}
M_\infty = \frac{V_{\infty}^*}{c_{\infty}^*}, \hspace{5mm}
Pr = \frac{\mu_{\infty}^* C_p}{k_{\infty}^*}
\end{eqnarray}
where $\mu_{\infty}^*$, $c_{\infty}^*$ and $k_{\infty}^*$ are freestream dynamic viscosity, speed of sound and thermal conductivity, respectively, and $C_p$ is the specific heat at constant pressure. For boundary layer flows over curved surfaces, we define the global G\"{o}rtler number as
$
G_{\lambda} = R_{\lambda}^2 \lambda^*/r^*
$,
where $r^*$ is the radius of the curvature.

\subsection{Compressible boundary-region equations}

 For a full compressible flow, the primitive form of the N-S equations in non-dimensional variables, are considered here in the form 

	\begin{eqnarray}\label{NS_1}
	\frac{D \bar{\rho}}{D t} 
	+ \rho \left( \frac{\partial \bar{u}}{\partial \bar{x}} + \frac{\partial \bar{v}}{\partial \bar{y}} + \frac{\partial \bar{w}}{\partial \bar{z}} \right) = 0
	\end{eqnarray}
	\begin{equation}\label{NS_2}
	\bar{\rho} \frac{D \bar{u}}{D \bar{t}}
	= -\frac{\partial \bar{p}}{\partial \bar{x}} 
	+ \frac{1}{Re_{\lambda}}\frac{\partial}{\partial \bar{x}} \left[ \frac{2}{3} \mu \left( 2\frac{\partial \bar{u}}{\partial \bar{x}} - \frac{\partial \bar{v}}{\partial \bar{y}} - \frac{\partial \bar{w}}{\partial \bar{z}} \right) \right]
	+ \frac{\partial}{\partial \bar{y}} \left[ \mu \left( \frac{\partial \bar{u}}{\partial \bar{y}} + \frac{\partial \bar{v}}{\partial \bar{x}} \right) \right]
	+ \frac{\partial}{\partial \bar{z}} \left[ \mu \left( \frac{\partial \bar{w}}{\partial \bar{x}} + \frac{\partial \bar{u}}{\partial \bar{z}} \right) \right]
	\end{equation}
	\begin{equation}\label{NS_3}
	\bar{\rho} \frac{D \bar{v}}{D \bar{t}}
	= -\frac{\partial \bar{p}}{\partial \bar{y}} 
	+ \frac{1}{Re_{\lambda}}\frac{\partial}{\partial \bar{y}} \left[ \frac{2}{3} \mu \left( 2\frac{\partial \bar{v}}{\partial \bar{y}} - \frac{\partial \bar{u}}{\partial \bar{x}} - \frac{\partial \bar{w}}{\partial \bar{z}} \right) \right]
	+ \frac{\partial}{\partial \bar{x}} \left[ \mu \left( \frac{\partial \bar{v}}{\partial \bar{x}} + \frac{\partial \bar{u}}{\partial \bar{y}} \right) \right]
	+ \frac{\partial}{\partial \bar{z}} \left[ \mu \left( \frac{\partial \bar{v}}{\partial \bar{z}} + \frac{\partial \bar{w}}{\partial \bar{y}} \right) \right]
	\end{equation}
	\begin{equation}\label{NS_4}
	\bar{\rho} \frac{D \bar{w}}{D \bar{t}}
	= -\frac{\partial \bar{p}}{\partial \bar{z}} 
	+ \frac{1}{Re_{\lambda}}\frac{\partial}{\partial \bar{z}} \left[ \frac{2}{3} \mu \left( 2\frac{\partial \bar{w}}{\partial \bar{z}} - \frac{\partial \bar{u}}{\partial \bar{x}} - \frac{\partial \bar{v}}{\partial \bar{y}} \right) \right]
	+ \frac{\partial}{\partial \bar{x}} \left[ \mu \left( \frac{\partial \bar{w}}{\partial \bar{x}} + \frac{\partial \bar{u}}{\partial \bar{z}} \right) \right]
	+ \frac{\partial}{\partial \bar{y}} \left[ \mu \left( \frac{\partial \bar{v}}{\partial \bar{z}} + \frac{\partial \bar{w}}{\partial \bar{y}} \right) \right]
	\end{equation}
	\begin{align}\label{NS_5}
	& \bar{\rho} \frac{D \bar{T}}{D \bar{t}}
	= 
	\frac{1}{Pr Re_{\lambda}} \left[ \frac{\partial}{\partial \bar{x}} \left( k \frac{\partial \bar{T}}{\partial \bar{x}} \right) + \frac{\partial}{\partial \bar{y}} \left( k \frac{\partial \bar{T}}{\partial \bar{y}} \right) + \frac{\partial}{\partial \bar{z}} \left( k \frac{\partial \bar{T}}{\partial \bar{z}} \right) \right]  \nonumber  \\
	&\quad - (\gamma - 1) M_{\infty}^2 \left[ p \left( \frac{\partial \bar{u}}{\partial \bar{x}} + \frac{\partial \bar{v}}{\partial \bar{y}} + \frac{\partial \bar{w}}{\partial \bar{z}} \right)
	-\frac{2}{3} \mu \left( \frac{\partial \bar{u}}{\partial \bar{x}} + \frac{\partial \bar{v}}{\partial \bar{y}} + \frac{\partial \bar{w}}{\partial \bar{z}} \right)^2 \right]
	+ (\gamma - 1) M_{\infty}^2 \frac{\mu}{Re_{\lambda}} \Psi 
	\end{align}
where
\begin{eqnarray}\label{Psi}
\Psi = 2\left( \frac{\partial \bar{u}}{\partial \bar{x}} \right)^2 + 2\left( \frac{\partial \bar{v}}{\partial \bar{y}} \right)^2 + 2\left( \frac{\partial \bar{w}}{\partial \bar{z}} \right)^2
+ \left( \frac{\partial \bar{u}}{\partial \bar{y}} + \frac{\partial \bar{v}}{\partial \bar{x}} \right)^2
+ \left( \frac{\partial \bar{w}}{\partial \bar{x}} + \frac{\partial \bar{u}}{\partial \bar{z}} \right)^2
+ \left( \frac{\partial \bar{v}}{\partial \bar{z}} + \frac{\partial \bar{w}}{\partial \bar{y}} \right)^2
\end{eqnarray}
is the dissipation function, and
$
D/D\bar{t} = \partial/\partial \bar{t} + \bar{u}\partial/\partial \bar{x} + \bar{v} \partial/\partial \bar{y} + \bar{w} \partial/\partial \bar{z}
$
is the substantial derivative (in what follows, we consider the steady-state case for the N-S equations where all time derivative terms drop out, i.e. $\partial / \partial \bar{t} = 0$). The pressure, $\bar{p}$, the temperature, $\bar{T}$  and the density, $\bar{\rho}$, of the fluid are combined in the equation of state in non-dimensional form, $\bar{p} = \bar{\rho} \bar{T} / \gamma M_{\infty}^2$, assuming non-chemically-reacting flows. The dynamic viscosity and thermal conductivity $k$ is linked to the temperature using the power law in dimensionless form:
$
\bar{\mu}= \bar{T}^b;  \hspace{1mm}
\bar{k} = C_p \bar{\mu}/Pr
$,
where $b=0.76$ (Ricco \& Wu \cite{Ricco}), $C_p = \gamma R / (\gamma - 1)$, $\gamma = 1.4$, and $Pr = 0.72$ for air.

In region III, $x/R_{\lambda}=O(1)$ implying that the streamwise ellipticity is weak and can be negligible (see Ricco \& Wu \cite{Ricco}, Ricco \cite{Ricco1}, or Marensi et al. \cite{Marensi}). 
Following this assumption, the streamwise distance can be re-scaled as $x = \bar{x}/R_{\lambda}$ ($x=O(1)$), whereas the cross-stream coordinates $(y,z)$ are fixed at $O(1)$, i.e $y = \bar{y}=O(1)$ and $z = \bar{z}=O(1)$. The temporal variation is also `re-scaled' as $t = \bar{t}/R_{\lambda}$. Finally, we take advantage of the crossflow components of velocity being asymptotically small compared to the streamwise component, and of the leading order pressure field being $O(R_{\lambda}^2)$, which suggests the asymptotic expansions with $1/R_{\lambda}$ as the small parameters:
\begin{eqnarray}\label{aa}
\bar{u}(x,y,z) = u(x,y,z) + ...; \hspace{4mm}  \bar{v}(x,y,z) = R_{\lambda}^{-1} v(x,y,z) + ...; \hspace{4mm}  \bar{w}(x,y,z) = R_{\lambda}^{-1} w(x,y,z) + ...;  \nonumber \\
\bar{\rho}(x,y,z) = \rho(x,y,z) + ...; \hspace{4mm}  \bar{p}(x,y,z) = R_{\lambda}^{-2} p(x,y,z) + ...; \bar{T}(x,y,z) = T(x,y,z) + ... 
\end{eqnarray}
Inserting \ref{aa} into the full Navier-Stokes equations in curvilinear coordinates, and retaining the first order terms in the expansions, we obtain the nonlinear compressible boundary region equations (NCBRE) in the form
%
	\begin{equation}\label{BRE_1}
		\textbf{V} \cdot \nabla \rho 
		+ \rho \nabla \cdot \textbf{V} = 0
	\end{equation}
	\begin{equation}\label{BRE_2}
		\rho \textbf{V} \cdot \nabla u
		= 
		\nabla_{c} \cdot \left( \mu \nabla_{c} u \right)
	\end{equation}
	\begin{equation}\label{BRE_3}
		\rho \textbf{V} \cdot \nabla v + G_\lambda u^2
		= 
		-\frac{\partial p}{\partial y} 
		+ \frac{\partial}{\partial y} \left[ \frac{2}{3} \mu \left( 3\frac{\partial v}{\partial y} - \nabla \cdot \textbf{V} \right) \right] 
		+ \frac{\partial}{\partial x} \left( \mu \frac{\partial u}{\partial y} \right) 
		+ \frac{\partial}{\partial z} \left[ \mu \left( \frac{\partial v}{\partial z} + \frac{\partial w}{\partial y} \right) \right]
	\end{equation}
	\begin{equation}\label{BRE_4}
		\rho \textbf{V} \cdot \nabla w
		= 
		-\frac{\partial p}{\partial z} 
		+ \frac{\partial}{\partial z} \left[ \frac{2}{3} \mu \left( 3\frac{\partial w}{\partial z} - \nabla \cdot \textbf{V} \right) \right]
		+ \frac{\partial}{\partial x} \left( \mu \frac{\partial u}{\partial z} \right)
		+ \frac{\partial}{\partial y} \left[ \mu \left( \frac{\partial v}{\partial z} + \frac{\partial w}{\partial y} \right) \right]
	\end{equation}
	\begin{equation}\label{BRE_5}
		\rho \textbf{V} \cdot \nabla T
		= 
		\frac{1}{Pr} \nabla_{c} \cdot \left( k \nabla_{c} T \right)
		+(\gamma - 1) M_{\infty}^2 \mu \left[ \left( \frac{\partial u}{\partial y} \right)^2
		+ \left( \frac{\partial u}{\partial z} \right)^2 \right]
	\end{equation}
	where $\textbf{V}$ is the velocity vector, $\nabla_{c}$ is the crossflow gradient operator, and $G_{\lambda} u^2$ is the term accounting for surface curvature.
	\begin{equation}\label{}
	\textbf{V} = u \vec{i} + v \vec{j} + w \vec{k}; \hspace{6mm} \nabla_{c} = \frac{\partial}{\partial y} \vec{j} + \frac{\partial}{\partial z} \vec{k}
	\end{equation}
This set of equations is parabolic in the streamwise direction and elliptic in the spanwise direction. Appropriate initial/upstream and boundary conditions are necessary to close the problem. The initial/upstream conditions could, for example, be the same as those used by Ricco \& Wu \cite{Ricco} (see also Ricco \cite{Ricco1}). However, for simplicity, in this work we excite the boundary layer using a small amplitude transpiration velocity ($v_w$) at the wall, in the form:
\begin{align}\label{dist}
v_w = A \sin \left[ \pi \frac{(x-x_s)}{(x_e-x_s)} \right]^2 \cos \left( \pi \frac{z}{\lambda} \right)
\end{align}
where $A$ is the amplitude, $x_s$ and $x_e$ are the start and end locations of the blowing and suction, respectively (in this study, $x_s=1.5\lambda^*$ and $x_e=4.5\lambda^*$), and $\lambda$ is the $O(1)$ spanwise separation. The NCBRE were numerically solved using an algorithm similar to the one employed by Sescu \& Thompson \cite{Sescu3} in the incompressible regime. We used second and fourth-order finite-difference schemes to, respectively, discretize the spatial derivatives in the wall-normal and spanwise directions. We employed periodic conditions in the spanwise direction to avoid compromising the stability and a staggered arrangement in the wall-normal direction to avoid decoupling between the velocity and pressure. We applied a first-order finite-difference marching scheme in the streamwise direction and converged the set of equations using a nonlinear time relaxation method. We found that the same numerical algorithm applied to the compressible regime is much faster since the continuity equation is no longer representing the divergence-free condition (making the set of equations stiff), but it is an equation for density. We placed the upper boundary far away from the wall ($10$ times the spanwise separation) and imposed vanishing gradients for all dependent variables.

The mean inflow condition is generated from a similarity solution obtained using the Dorodnitsyn-Howarth coordinate transformation
$
\bar{Y}(x,y) = \int_{0}^{y} \rho(x,\tilde{y}) d \tilde{y}
$.
A similarity variable is defined as
$
\eta = \bar{Y}\left( Re_{x}/2x \right)^{1/2},
$
where $Re_{x}$ is the Reynolds number calculated based on the freestream velocity and the distance from the leading edge. The base velocity and temperature can be expressed as
\begin{equation}\label{ssd}
U = F'(\eta), \hspace{4mm} V = (2 x Re_{x})^{-1/2} (\eta_c T F' - T F), \hspace{4mm} T = T(\eta)
\end{equation}
where prime represents differentiation with respect to $\eta$, and $\eta_c = 1/T \int_{0}^{\eta} T(\tilde{\eta}) d \tilde{\eta}$. $F$ and $T$ satisfy the following coupled equations
\begin{eqnarray}\label{bl}
&& \left( \frac{\mu}{T} F'' \right)' + F F'' = 0, \nonumber \\
&& \frac{1}{Pr} \left( \frac{\mu}{T} T' \right)' + F T' + (\gamma - 1) M^2 \frac{\mu}{T} F''^2 = 0,
\end{eqnarray}
subjected to the boundary conditions
$
F(0)=F'(0)=0, \hspace{4mm} T'(0)=0, \hspace{4mm} F' \rightarrow 1, \hspace{4mm} T \rightarrow 1 \rightarrow
\hspace{2mm} as \hspace{2mm}  \eta \rightarrow \infty
$.Equations (\ref{bl}) were solved numerically to determine $F$ and $T$ and then used in equations (\ref{ssd}) to obtained the mean inflow condition.


\section{Results and Discussion}\label{}

\begin{figure}[H]
 \begin{center}
   \includegraphics[width=15.cm]{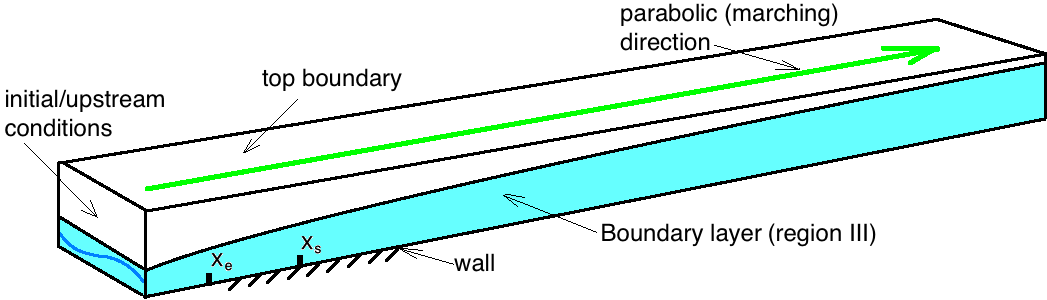} 
 \end{center}
  \caption{\label{} Flow domain.}
   \label{f3}
\end{figure}

We consider G\"{o}rtler vortices developing in high-speed boundary layers, with the Mach number ranging from supersonic, $M_\infty=2$, to hypersonic conditions, $M_\infty=6$. We neglect chemical reactions inside the boundary layer at hypersonic speeds. We vary the spanwise separation of the vortices between $0.3$ cm and $0.7$ cm. We calculate the similarity solution imposed at the upstream boundary using the Reynolds number based on the distance from the leading edge, which is maintained constant for a given Mach number (note that the Reynolds number based on the spanwise separation differs for each $\lambda$). The radius of curvature of the concave surface is $1.6$ m for all cases involved in the parametric study. The boundary layer flow is excited by a small disturbance applied to the vertical velocity imposed at the wall, with an amplitude of $0.2$\% of the freestream velocity (the form of this disturbance is given in equation (\ref{dist})). Figure \ref{f3} shows a sketch of the flow domain, with region III highlighted in blue. We impose the similarity type velocity and temperature profiles for a compressible boundary layer at the upstream boundary, located on the left hand side of the sketch. At the wall, we impose the no-slip boundary condition for the velocity and either an isothermal or adiabatic boundary condition for the temperature field. We impose vanishing gradients at the top boundary, and a symmetry condition along the spanwise direction since we only simulate one streamwise vortex (corresponding to half of the mushroom shape), belonging to the pair of counter-rotating vortices.

\begin{figure}[H]
 \begin{center}
   \includegraphics[width=7.cm]{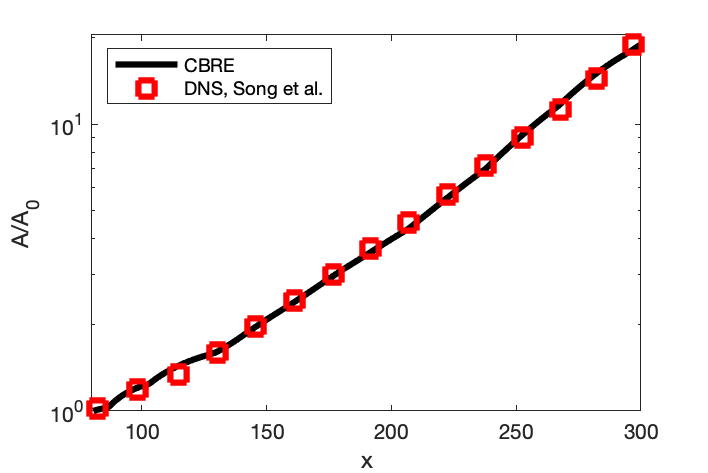}
 \end{center}
  \caption{Comparison to DNS results of Song et al. \cite{Song}.}
  \label{f4}
\end{figure}

We validate the accuracy of our method by comparing our results to the DNS results of Song et al. \cite{Song}. We use a disturbance similar to that employed in their study with an amplitude of $0.5$\% for the transpiration velocity; note, however, that they imposed the disturbance at the inlet boundary, whereas in our simulation, we impose it at the wall. In Song et al., the freestream Mach number is $6.5$ and the radius of curvature is $1.6$ m, the same as in our parametric study. In figure \ref{f4}, we plot the scaled amplitude, $A = \max_{y,z}(T')$, based on the temperature disturbance calculated from each $(y,z)$-plane at every fixed streamwise location ($A_0$ in figure \ref{f4} is the same amplitude calculated at the $x$-coordinate of the imposed disturbance). Our results in figure \ref{f4} compare very well with the DNS in Song et al. (note that the streamwise coordinate has been scaled properly to match the DNS range).

\begin{figure}[H]
 \begin{center}
   (a) \includegraphics[width=11.cm]{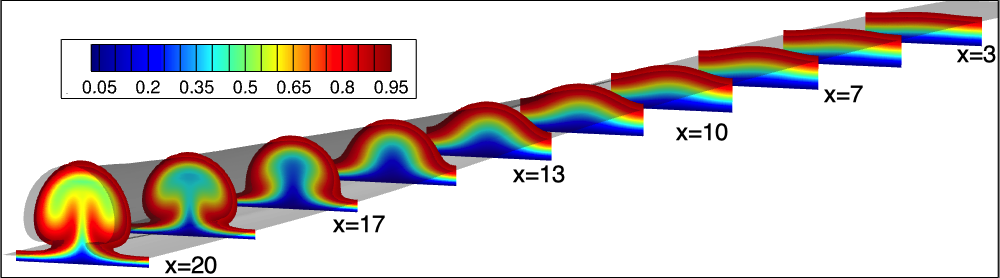}\\
   (b) \includegraphics[width=11.cm]{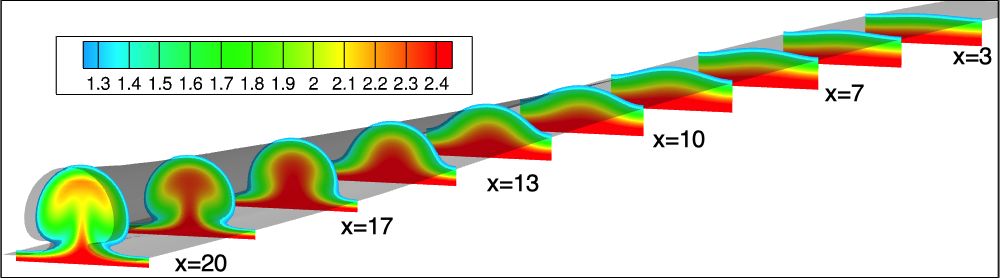}
 \end{center}
  \caption{Contour plots of velocity magnitude (a) and temperature (b) in crossflow planes  for $M_\infty=3$, $\lambda=0.4$, and adiabatic wall condition.}
  \label{f5}
\end{figure}

\begin{figure}[H]
 \begin{center}
  (a) \includegraphics[width=11.cm]{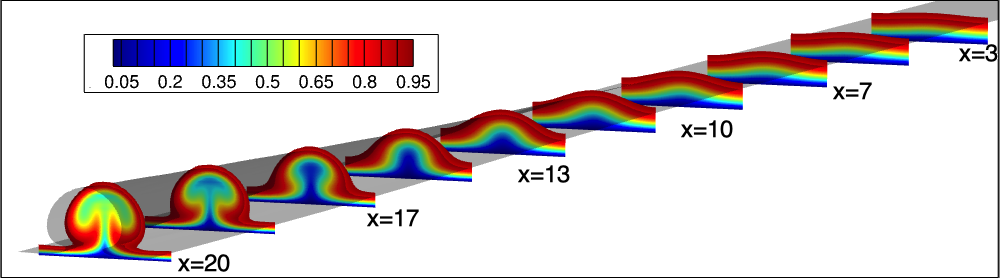}\\
  (b) \includegraphics[width=11.cm]{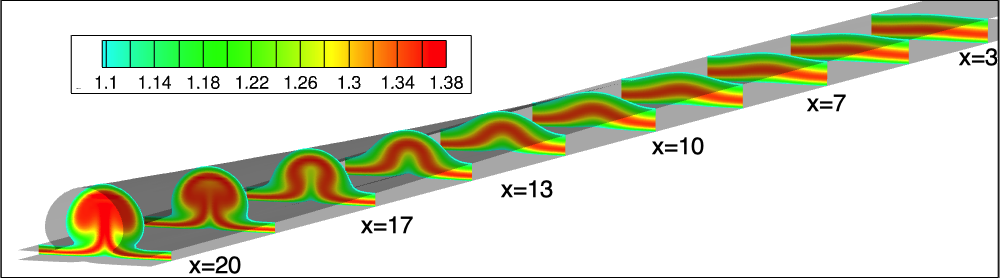}
   \end{center}
   \caption{Contour plots of velocity magnitude (a) and temperature (b) in crossflow planes for $M_\infty=3$, $\lambda=0.4$, and isothermal wall condition.}
  \label{f6}
\end{figure}

We focus on the parametric study by varying the Mach number and the spanwise separation. In figures \ref{f5} and \ref{f6}, we show contour plots of temperature in consecutive cross-stream planes, for a Mach number $M_\infty=3$ and a spanwise separation $\lambda=0.4$ cm. The contour plots in figure \ref{f5} correspond to the adiabatic wall-condition, while those in figure \ref{f6} to the isothermal condition. They both illustrate the streamwise development of the velocity magnitude and the temperature field of the G\"{o}rtler vortices, which display the mushroom-shape characteristic structures. For both wall conditions, blue and green colors in the velocity magnitude contours correspond to the low-speed streaks, while the red color regions are associated with high-speed streaks. The temperature contour plots of the adiabatic wall-condition indicate that the low-speed streaks are associated with an increase in temperature (in red or yellow). The high-speed streaks are characterized by a lower temperature level (shown here in green). For the isothermal wall-condition case, however, it appears as if the high temperature region of the structures is sandwiched between two cooler regions belonging to both low and high speed streaks. This `sandwich' phenomenon leads to a major difference between the two wall conditions in terms of the height that these mushroom-shape structures take. In the adiabatic wall-condition case, for example, a significant increase in temperature at the wall generates more convection, which contributes to the growth of `thermals' inside the structures. In contrast, the cooler regions at the wall result in less convection and therefore smaller vortex (i.e mushroom) structures in the isothermal case.

\begin{figure}[H]
 \begin{center}
 \includegraphics[width=4.0cm]{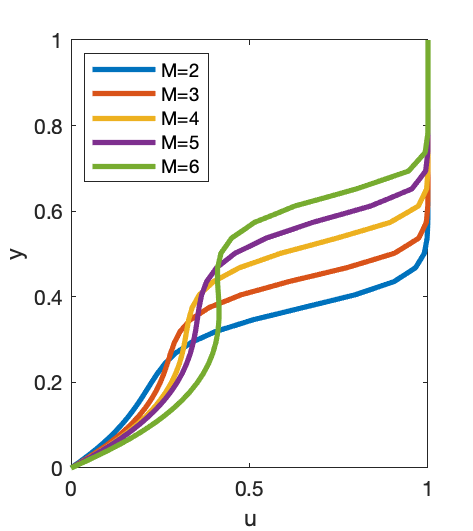} 
 \includegraphics[width=4.0cm]{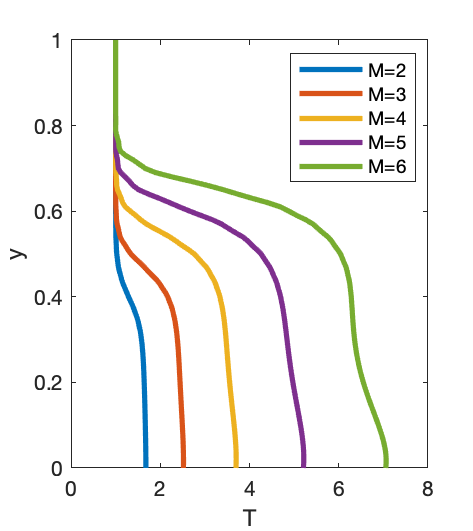} 
 \includegraphics[width=4.0cm]{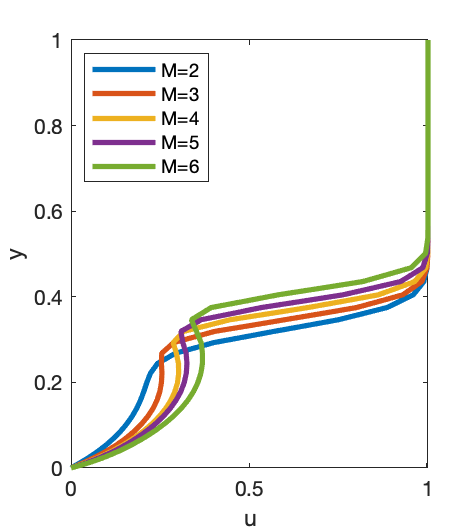}
 \includegraphics[width=4.0cm]{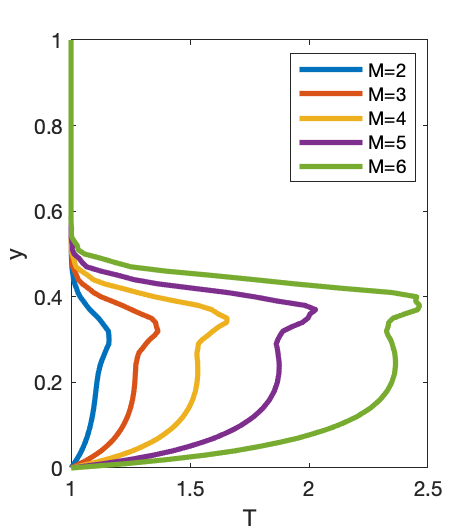}  \\
 \hspace{3mm} (a) \hspace{34mm} (b) \hspace{34mm} (c) \hspace{34mm} (d)
 \end{center}
 \vspace{-7mm}
  \caption{\label{} Velocity and temperature boundary layer profiles in $z=0$, for different Mach numbers: (a) velocity, adiabatic wall; (b) temperature, adiabatic wall; (c) velocity, isothermal wall; (d) temperature, isothermal wall.}
  \label{f7}
\end{figure}

\begin{figure}[H]
 \begin{center}
 \includegraphics[width=4.0cm]{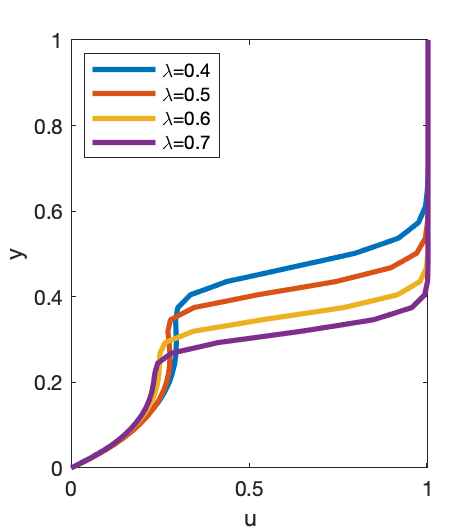} 
 \includegraphics[width=4.0cm]{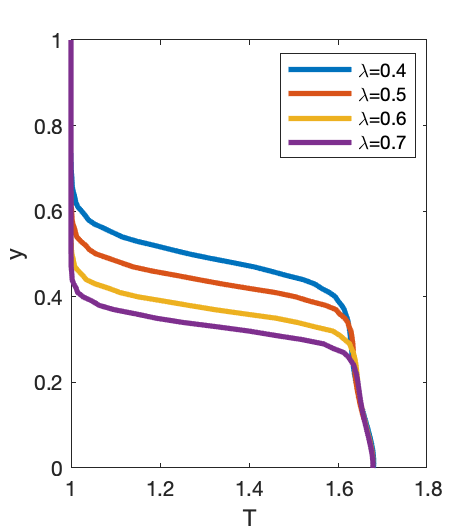} 
 \includegraphics[width=4.0cm]{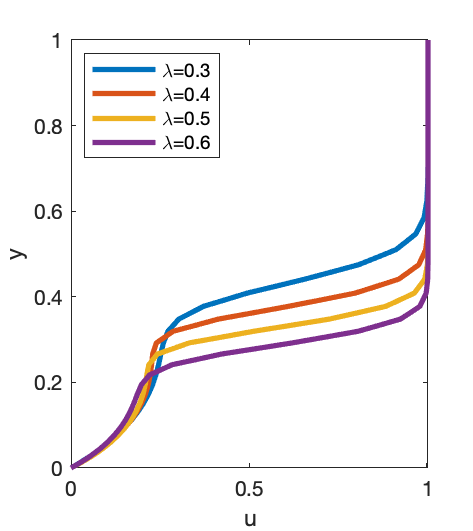}
 \includegraphics[width=4.0cm]{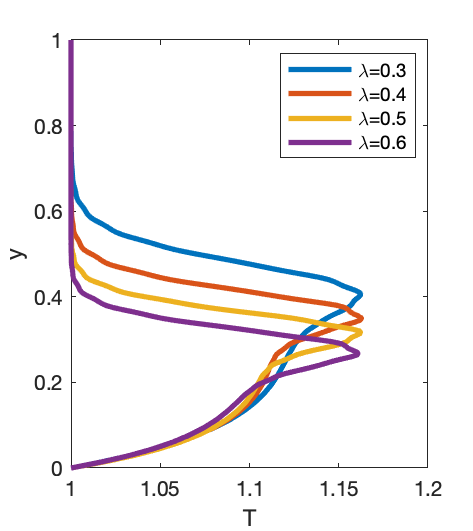}  \\
 \hspace{3mm} (a) \hspace{34mm} (b) \hspace{34mm} (c) \hspace{34mm} (d)
 \end{center}
 \vspace{-7mm}
  \caption{\label{} Velocity and temperature boundary layer profiles in $z=0$, for different spanwise separations: a) velocity, adiabatic wall; b) temperature, adiabatic wall; c) velocity, isothermal wall; d) temperature, isothermal wall.}
  \label{f8}
\end{figure}

\begin{figure}[H]
 \begin{center}
 \includegraphics[width=4.9cm]{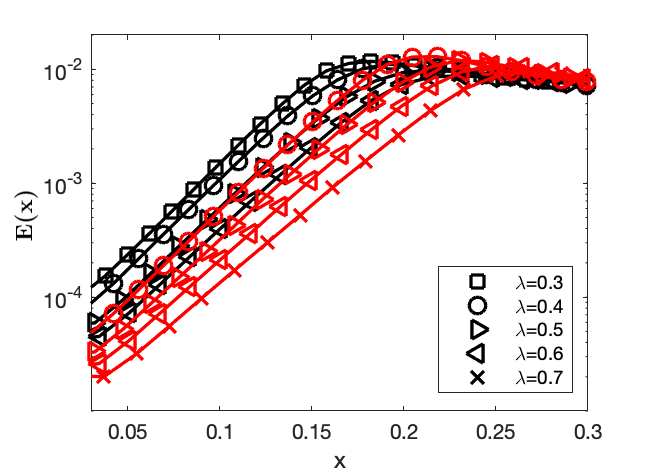} 
 \includegraphics[width=4.9cm]{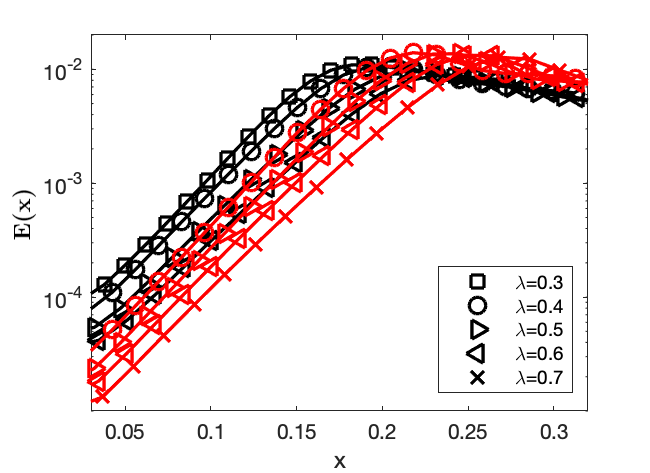} 
 \includegraphics[width=4.9cm]{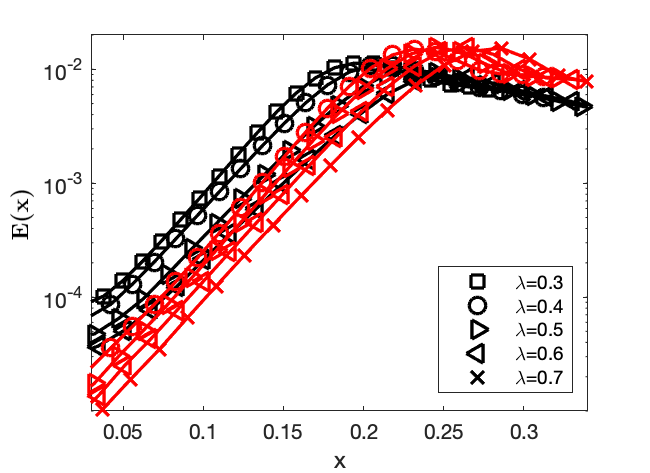}  \\ 
 \hspace{2mm} (a) \hspace{4.35cm} (b) \hspace{4.3cm} (c) \\
 \includegraphics[width=4.9cm]{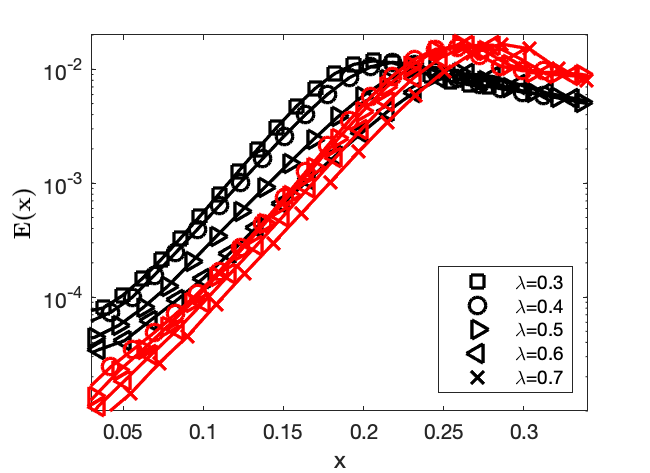}  
 \includegraphics[width=4.9cm]{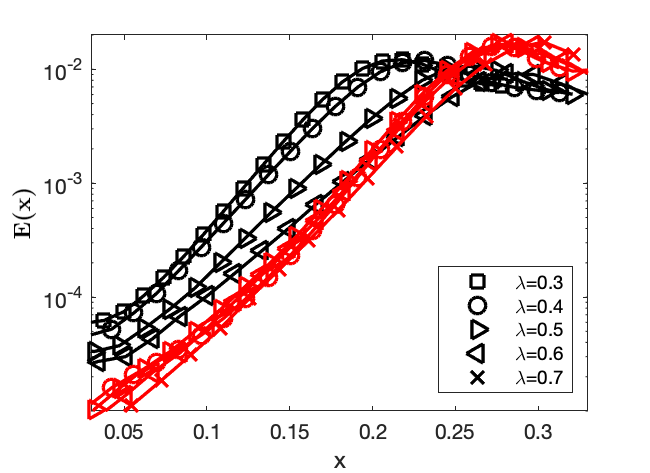}  \\
 \hspace{2mm} (d) \hspace{4.3cm} (e)
 \end{center}
 \vspace{-7mm}
  \caption{\label{} Vortex energy distribution along the streamwise direction for both isothermal (in black) and adiabatic (in red) wall conditions: a) $M_\infty=2.0$; b) $M_\infty=3.0$; c) $M_\infty=4.0$; d) $M_\infty=5.0$; e) $M_\infty=6.0$}
  \label{f9}
\end{figure}

Figures \ref{f7} and \ref{f8} show selected velocity and temperature profiles along the wall-normal direction at the spanwise location corresponding to the center of the mushroom-shape structures and the streamwise location corresponding to the highest vortex energy (see figure \ref{f9}). In figure \ref{f7}, we plot the profiles at different Mach numbers (keeping the spanwise separation constant at $\lambda=0.4$ cm), whereas in figure \ref{f8}, we repeat this plot at different spanwise separations (keeping the Mach number constant at $M_\infty=2$). As we increase the Mach number, the `height' of the streamwise vortices (i.e the size of the mushroom-shape structures) increases (figs. \ref{f7}a and \ref{f7}c) as a result of the increase in the temperature close to the wall (figs. \ref{f7}b and \ref{f7}d). The increase in size is more evident in the adiabatic case, confirming the aforementioned observations. For the highest Mach number ($M_\infty=6$) the temperature near the wall increases by a factor of $6$ for the adiabatic case and $2.5$ in isothermal conditions, compared to the lowest considered Mach number of $M_\infty=2$. It is evident from figure \ref{f8} that as we increase the spanwise separation, the vertical size of the mushroom-shape structures decreases (almost linearly in fact) for both streamwise velocity and temperature fields. 

We quantify the vortex energy as
\begin{equation}\label{jj}
E(x) = \intop_{z_1}^{z_2}  \intop_{0}^{\infty}  \left[ \left| u(x,y,z) - u_m(x,y) \right|^{2} +  \left| v(x,y,z) - v_m(x,y) \right|^{2} +  \left| w(x,y,z) - w_m(x,y) \right|^{2} \right] dzdy,
\end{equation}
where $u_m(x,y)$, $v_m(x,y)$, and $w_m(x,y)$ are the spanwise mean components of velocity, and $z_1$ and $z_2$ are the coordinates of the spanwise domain boundaries.

In figure \ref{f9}, we plot the vortex energy from equation $(13)$ against the streamwise coordinate, $x$. For each sub-figure, we fix the Mach number at a constrant value and vary the spanwise separation, $\lambda$. At all Mach numbers we find that, as we increase the spanwise separation, there is a noticeable reduction in the scaled energy for both isothermal (in black) and adiabatic (in red) wall conditions; the energy saturation location moves downstream as the spanwise separation increases. 
The drop in the vortex energy associated with the increase in the spanwise separation becomes less evident as the Mach number increases, particularly in the adiabatic wall-condition as the curves of the different $\lambda$ values coalesce until they almost fall on top of each other at the highest Mach number (figure \ref{f9}e). One possible explanation of this observation is that the reduction in the vortex energy caused by increasing $\lambda$ diminishes when the convection levels are increased at the wall, particularly for high Mach numbers (note that as the Mach number increases, the temperature in proximity to the wall increases considerably; see figure \ref{f7}b). Another interesting aspect that we can extract from figure \ref{f9} is the effect that an adiabatic wall condition has on the vortex energy development: i.e., it appears that the energy growth is delayed in the adiabatic case, while the peak energy appears to be fixed at the level for the isothermal case in relatively low supersonic Mach number conditions (figures \ref{f9}a and \ref{f9}b). On the other hand, this level is slightly higher at high supersonic or hypersonic flow conditions (figures \ref{f9}c, \ref{f9}d and \ref{f9}e).

\begin{figure}[H]
 \begin{center}
 \includegraphics[width=4.9cm]{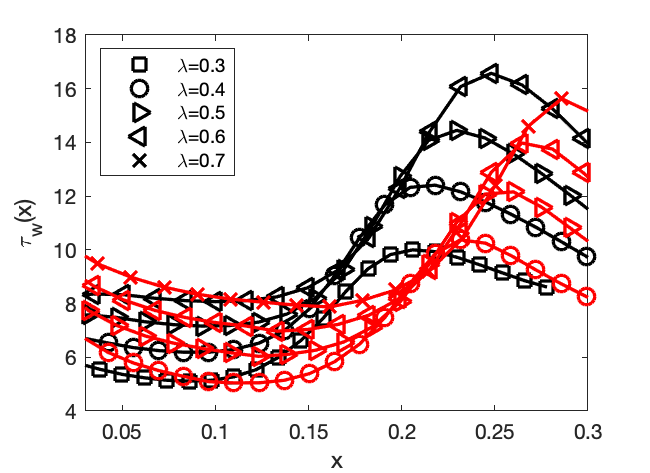} 
 \includegraphics[width=4.9cm]{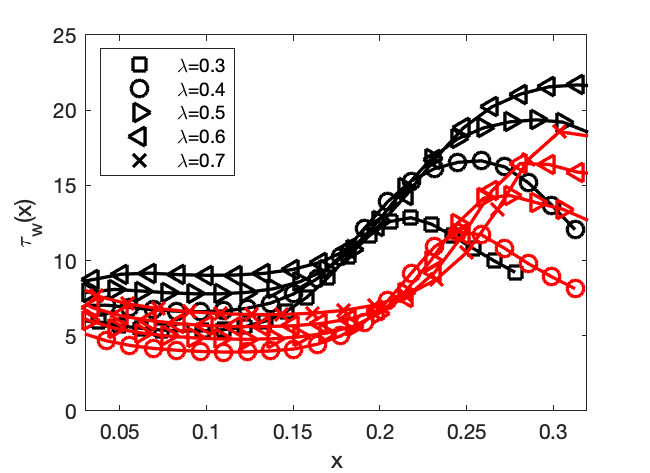} 
 \includegraphics[width=4.9cm]{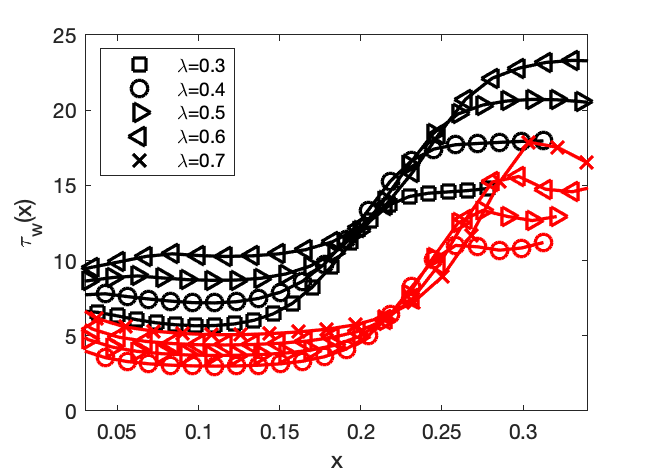}  \\ 
 \hspace{2mm} (a) \hspace{4.35cm} (b) \hspace{4.3cm} (c) \\
 \includegraphics[width=4.9cm]{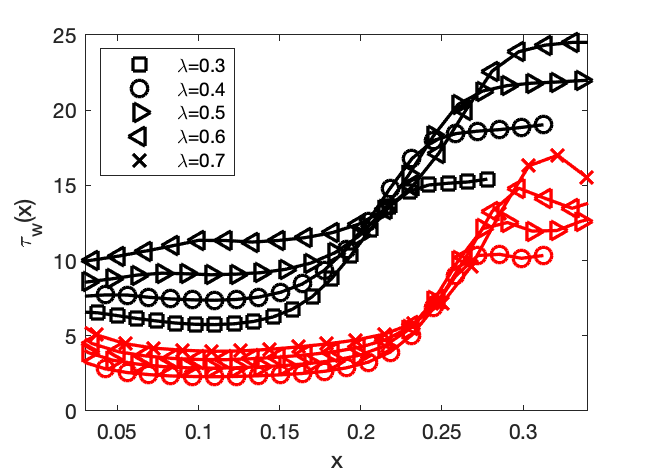}  
 \includegraphics[width=4.9cm]{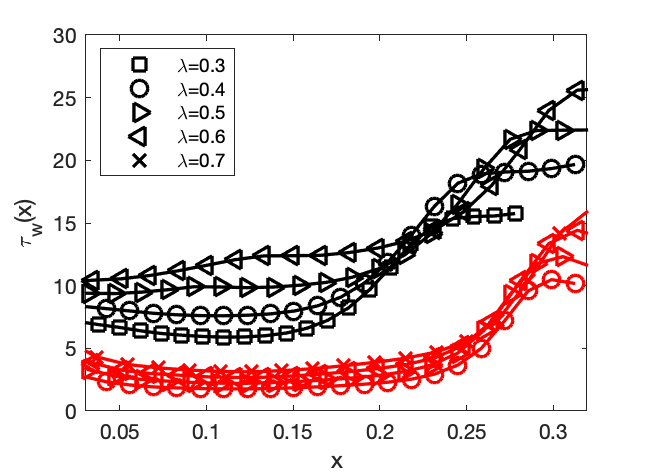}  \\
 \hspace{2mm} (d) \hspace{4.3cm} (e)
 \end{center}
  \caption{\label{} Spanwise averaged wall shear stress distribution along the streamwise direction: a) $M_\infty=2.0$; b) $M_\infty=3.0$; c) $M_\infty=4.0$; d) $M_\infty=5.0$; e) $M_\infty=6.0$; isothermal (in black) and adiabatic (in red) wall conditions}
  \label{f10}
\end{figure}

The spanwise averaged wall shear stress evaluated using the integral

\begin{equation}\label{wall shear stress}
\tau_w(x) =  \frac{1}{(z_2 - z_1)}\intop_{z_1}^{z_2}  \left.  \frac{\partial u}{\partial y} \right|_{y=0} (x,0,z)  dz
\end{equation} 
is plotted in figure \ref{f10} against the streamwise coordinate for both isothermal and adiabatic conditions. For each case, we fix the Mach number at a constant value and vary the spanwise separation. The results show that, in all considered cases, the wall shear stress increases as the spanwise separation increases. Also, as expected, the wall shear stress of the adiabatic wall-condition is lower than that of the isothermal case as a result of the high level of heating in proximity to the wall (this was observed in previous studies, such as Spall and Malik \cite{Spall}, Elliot \cite{Elliot}, and Sescu et al. \cite{Sescu4} etc.). The jump in the wall shear stress coincides (approximately) with the same location associated with the the energy saturation initiation point. The effect of increasing $\lambda$ on the shear stress becomes less apparent in the case of the highest Mach number (i.e $M_\infty=6$) in adiabatic wall-conditions. 

\begin{figure}[H]
 \begin{center}
 \includegraphics[width=4.9cm]{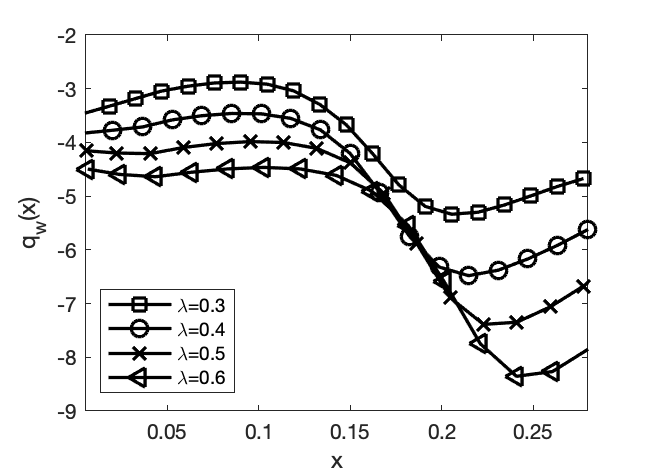} 
 \includegraphics[width=4.9cm]{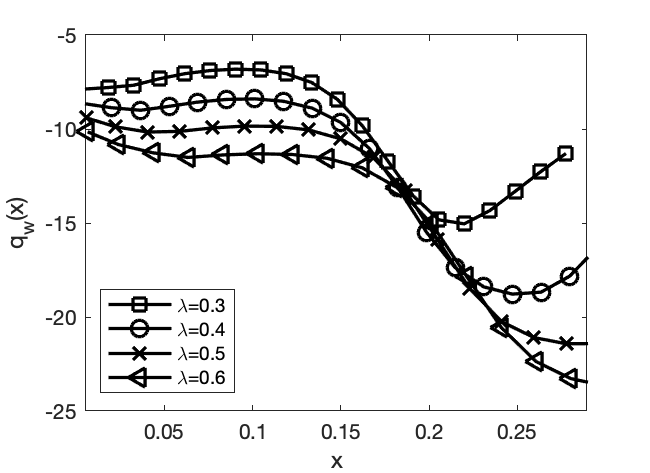} 
 \includegraphics[width=4.9cm]{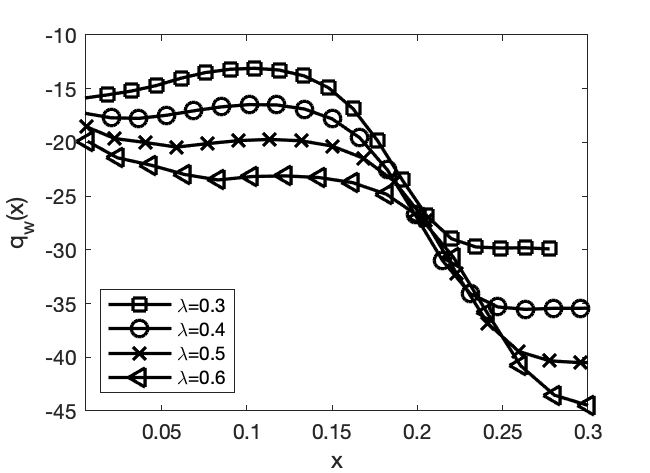}  \\ 
 \hspace{2mm} (a) \hspace{4.35cm} (b) \hspace{4.3cm} (c) \\
 \includegraphics[width=4.9cm]{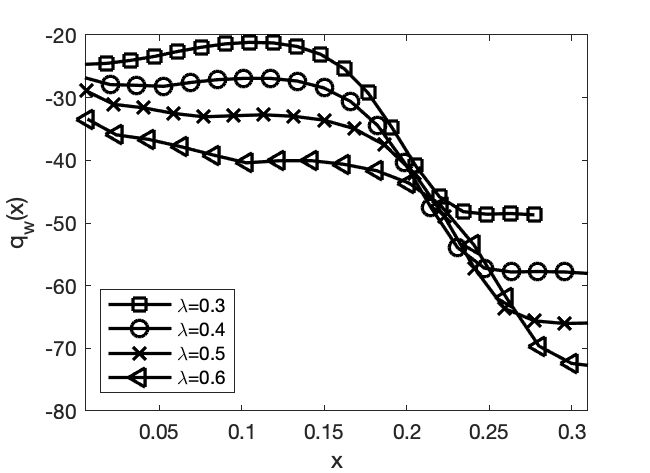}  
 \includegraphics[width=4.9cm]{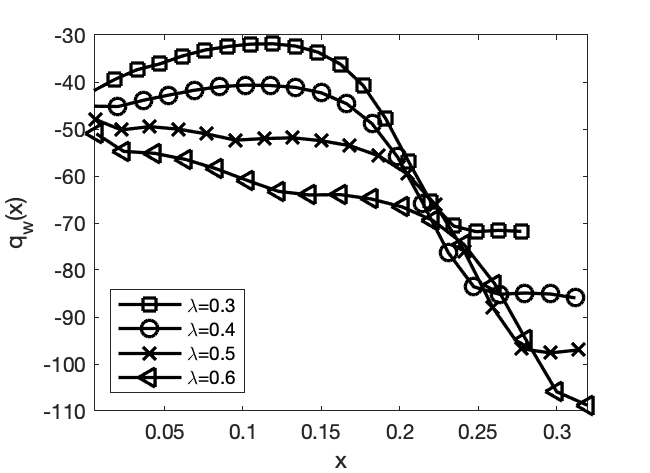}  \\
 \hspace{2mm} (d) \hspace{4.3cm} (e)
 \end{center}
  \caption{\label{} Spanwise averaged wall heat flux distribution along the streamwise direction: a) $M_\infty=2.0$; b) $M_\infty=3.0$; c) $M_\infty=4.0$; d) $M_\infty=5.0$; e) $M_\infty=6.0$; isothermal (in black) and adiabatic (in red) wall conditions}
  \label{f11}
\end{figure}

The spanwise averaged wall heat flux is calculated according to
\begin{equation}\label{jj}
q_w(x) = - \frac{1}{(z_2 - z_1)}\intop_{z_1}^{z_2}  \left.  \frac{\partial T}{\partial y} \right|_{y=0} (x,0,z)  dz,
\end{equation}
which is plotted in figure \ref{f11} for the isothermal wall-condition (the wall heat flux is zero for the adiabatic counterpart) as a function of the streamwise coordinate. It is evident from this figure that the wall heat flux decreases as the spanwise separation is increased. Moreover, we notice a decay of the wall heat flux in the streamwise coordinate range where the energy saturation takes place. 

The parabolic character of the NCBRE framework allows the solution to be determined efficiently by a marching procedure in the streamwise direction. This makes the numerical algorithm very fast, proving it suitable for parametric studies that can be conducted in a timely manner. In table \ref{t1}, we show the CPU time for different cases for a grid resolution of $600$ points in the streamwise direction, $201$ points along $y$, and $81$ points along $z$; the spanwise length is limited to the spanwise separation of the vortices, while the in the wall normal direction the domain size is $5$ times the spanwise separation, with the grid stretched toward the upper boundary. In addition, the method does not require large computational resources (in this paper the simulations were all run on a laptop computer with Intel Core i7 processor). 

\begin{table}[H]
\caption{CPU time for different cases}
\centering 
\begin{tabular}{c c c } 
\\\hline 
Mach number & CPU time (isothermal) & CPU time (adiabatic) \\ [0.5ex] 
\hline 
2.0 & 1.9 min & 2.1 min \\ 
3.0 & 2.2 min & 2.3 min \\
4.0 & 2.6 min & 2.7 min \\
5.0 & 2.7 min & 2.9 min \\
6.0 & 2.9 min & 3.1 min \\ [1ex] 
\hline 
\end{tabular}
\label{t1} 
\end{table}

\section{Conclusions}

In this paper, G\"{o}rtler vortices that develop in high-speed boundary layer flows over concave surfaces are investigated using a numerical solution to the nonlinear compressible boundary region equations (NCBRE). We targeted the nonlinear development of the centrifugal instabilities that develop on the surface, by varying the spanwise separation that of the upstream disturbance (this dictates the spanwise separation of the downstream G\"{o}rtler vortices) and upstream inflow the Mach number. The boundary layer was excited using a small disturbance at the wall in the form of steady blowing and suction.

We considered a wide range of spanwise separations and the Mach numbers covering both supersonic and hypersonic regimes. Contours of velocity and temperature at various crossflow planes showed the vortex development in the form of mushroom shape structures evolving in the streamwise direction. The kinetic energy of the primary instability was calculated and plotted against the streamwise coordinate. It was observed that as the upstream spanwise separation is increased, the scaled kinetic energy maximum increases, and that the streamwise location where the energy saturation sets in moves further downstream. We also calculated the wall shear stress and the wall heat flux and observed that the wall shear stress increases as the spanwise separation increases, and - as expected - the wall shear stress is smaller for the adiabatic condition, as a result of the significant increase in temperature in proximity to the wall. There was also a jump of the wall shear stress approximately at the streamwise location corresponding to the point of energy saturation initiation. For the isothermal wall condition, the wall heat flux showed a characteristic decrease as the spanwise vortex separation is increased; there was also decay of the wall heat flux in this case at the streamwise location corresponding to the point where the energy saturation is initiated. 

The framework of the NCBRE mathematical model is robust and the associated numerical algorithm provides results in a very short time compared to other mathematical models such as DNS or PSE, for example. The simulations were run on a single processor, which makes our BRE approach suitable for multiple parametric studies.



\end{document}